\documentclass [10pt]{article}
\usepackage{amsmath}
\usepackage{amsfonts}
\usepackage{amssymb}
\makeatletter \oddsidemargin  0in \evensidemargin 0in
\textwidth16cm \RequirePackage[dvips]{graphicx} \textheight 20.5cm
\newcommand{\singlespacing}{\let\CS=\@currsize\renewcommand{\baselinestretch}{1.5}\tiny\CS}
\makeatletter \oddsidemargin  -.1in \evensidemargin -.1in
\newcommand{\doublespacing}{\let\CS=\@currsize\renewcommand{\baselinestretch}{1.35}\tiny\CS}

\def\@citex[#1]#2{\if@filesw\immediate\write\@auxout{\string\citation{#2}}\fi
  \def\@citea{}\@cite{\@for\@citeb:=#2\do
    {\@citea\def\@citea{,\linebreak[0]\hskip0pt plus .2em}%
      \@ifundefined{b@\@citeb}%
    {{\bf ?}\@warning{Citation `\@citeb' on page \thepage\space undefined}}%
      \hbox{\csname b@\@citeb\endcsname}}}{#1}}

\newtheorem{rule-def}[theorem]{Rule}

\begin{document}

\title{\bf Common entanglement witnesses and their characteristics}
\author{ Nirman Ganguly$^{1,3}$\thanks{Corresponding Author:nirmanganguly@gmail.com},~~Satyabrata Adhikari$^{2}$,~~A. S. Majumdar$^{3}$\\
$^1$ Dept. of Mathematics, Heritage Institute of Technology, Kolkata-107, West Bengal, India\\
$^2$ Institute of Physics, Sainik School Post, Bhubaneshwar-751005, Orissa, India \\
$^3$ S. N.Bose National Centre for Basic Sciences, Salt Lake,
Kolkata-700 098, India\\}
\maketitle{}

\begin{abstract}
We investigate the issue of finding common entanglement witness for
certain class of states and extend this study to the case of 
Schmidt number witnesses. We also introduce the notion of common 
decomposable and non-decomposable witness operators which is specially 
useful for constructing a common witness where one of the entangled states 
is with a positive partial transpose. Our approach is illustrated with the help 
of suitable examples of qutrit systems.
\end{abstract}

\section{Introduction }

Entanglement lies at the core of quantum information theory.
Although entanglement is a vital resource for processes like teleportation, 
dense coding, quantum key distribution and quantum computation
\cite{1,2,3,4}, its detection is a hard task.
For low dimensional ($2\otimes2$ and $2\otimes3$) states there
exists a simple necessary and sufficient condition for separability
\cite{5,6}, which is based on the fact that separable states have
a positive partial transpose (PPT). For higher dimensional systems
all states with negative partial transpose (NPT) are entangled but
there are entangled states which have a positive partial
transpose. This paradoxical behaviour of quantum entanglement in
higher dimensions makes it difficult to lay down a single
necessary and sufficient condition for its detection.

A major breakthrough in this direction is provided in the form of entanglement
witnesses (EW) \cite{6,7}. An outcome of the celebrated Hahn-Banach
theorem in functional analysis, entanglement witnesses are
hermitian operators with at least one negative eigenvalue.
Entanglement witness provides a necessary and sufficient condition
to detect entanglement. More specifically a given state is
entangled if and only if there is an EW that detects it \cite{6}.
Such a property makes EW an useful tool in the 
experimental detection of entanglement. 
Though it is difficult to construct an EW that detects an unknown entangled
state, several methods have been suggested in the literature pertaining
to several classes of entangled states\cite{7,8,9}.
The notion of an entanglement witness was further extended to Schmidt number 
witness, which detects the Schmidt number of quantum states \cite{10,11}.

An interesting line of study concerning entanglement witnesses is to find 
a common EW for different entangled states.  It was proved by Wu and 
Guo \cite{12}  that 
for a given pair of entangled states $\rho_{1}$ and $\rho_{2}$, a common
EW exists if and only if $ \lambda\rho_{1}+(1-\lambda)\rho_{2} $ is an entangled
state $\forall \lambda \in [0,1]$. They thus arrived at a sufficient condition 
for entanglement for pairs of entangled states. Construction of a common
entanglement witness for two entangled states not only detects
them but also any state which is a convex combination of the two.
Thereby one can detect a large class of entangled states if one is able
to find a common EW satisfying the above criterion.

In the present work our motivation is to propose some methods to construct 
common EW for certain classes of states making use of the above condition of
existence. We first propose some characteristics of common Schmidt 
number witnesses based on the analysis of common entanglement witnesses, 
providing suitable examples for our propositions. We then
suggest schemes for finding common EW for various categories of states based 
on their spectral characteristics. The distinction between  a common 
decomposable witness operator and a non-decomposable one is of relevance in
the probe for finding common EW. A decomposable operator is unable to detect 
a PPTES (positive partially transposed
entangled state), whereas a non-decomposable witness can successfully detect a
PPTES. This distinction propagates to a common witness. Precisely, if one
of the entangled states in a convex combination is a PPTES, then the common 
witness is non-decomposable. Our analysis makes use of some decomposable and 
nondecomposable witnesses including one which we had proposed earlier \cite{14}.
We illustrate our results through various appropriate examples
from qutrit systems.

The paper is organized as follows. 
We begin  with a statement of certain relevant
definitions and results in section 2, which are useful for the later analysis. 
In section 3, we propose and study some features of common Schmidt number 
witnesses. Next, in section 4 we suggest methods to detect a combined pair of 
entangled states and construct the common EW for them. 
We then provide explicit examples demonstrating our methods for finding common 
entanglement witnesses in section 5. In section 6, we distinguish between a 
common decomposable and a  non-decomposable witness operator  citing examples. 
We conclude with a brief summary of our results in section 7.

\section{Prerequisites}

We begin with a brief summary of a few useful definitions and results.\\
\textbf{Definition-1:} The kernel of a given density matrix $\rho
\in B(H_{A}\otimes H_{B})$ is defined as
\begin{eqnarray}
ker(\rho)=\{|x\rangle\in H_{A}:\rho|x\rangle=0\}
\end{eqnarray}
\textbf{Definition-2:} A PPT entangled state $\delta$ is called an
edge state if for any $\varepsilon>0$ and any product vector
$|e,f\rangle$, $\delta^{'}=\delta-\varepsilon|e,f\rangle\langle
e,f|$ is not a PPT state \cite{8}.\vskip0.2cm

\noindent \textbf{Definition-3:} A hermitian operator $W$ is said
to be an entanglement witness operator iff
\begin{eqnarray}
&&(i)~ Tr(W\sigma)\geq 0~~~  \forall~~  \textrm{separable state}~
\sigma~~ and {}\nonumber\\&& (ii)~Tr(W\rho)<0 ,~~ \textrm{for at
least one entangled state}~\rho. \label{def.3}
\end{eqnarray}
\textbf{Definition-4:} A witness operator is said to be
decomposable if it can be expressed in the form
\begin{eqnarray}
D= P+Q^{T_{A}}\label{decomwit.}
\end{eqnarray}
where $P$ and $Q$ are positive semi-definite operators.
Non-decomposable operators are those which cannot be written as in
(\ref{decomwit.}).\\
 \textbf{Result-1:} A non-decomposable
witness that can detect an edge state $\delta$ is of the form $
P+Q^{T_{A}}-\varepsilon I $, where $P$ is a projector on
ker($\delta$) and $Q$ a projector on ker($\delta^{T_{A}}$) and
$0<\varepsilon\leq\varepsilon_{0}=inf_{|e,f\rangle}\langle
e,f|P+Q^{T_{A}}|e,f\rangle$~~($|e,f\rangle$ is a product vector)
\cite{8}.\vskip0.2cm

\noindent \textbf{Result-2:} Another non-decomposable witness
operator can be expressed in the form as \cite{14}:
\begin{eqnarray}
W_{1}=Q^{T_{A}}-k(I-P), \label{nondecompowitness}
\end{eqnarray}
where $0<k\leq k_{0}$ and
\begin{eqnarray}
k_{0}= \textrm{min}
\frac{Tr(Q^{T_{A}}\sigma)}{Tr((I-P)\sigma)}\label{k0}
\end{eqnarray}
The minimum is taken over all separable states
$\sigma$. An extension of these in tripartite systems is the 
witness 
\begin{eqnarray}
W_{tri}=Q^{T_{X}}-k_{0}(I-P),~~~~X=1,2,3 \label{multiwitness}
\end{eqnarray}
$P$=Projector on Ker($\delta_{tri}$) and $Q$= Projector on
Ker($\delta_{tri}^{T_{X}}$),
 where $T_{X}$ denotes the transpose taken with respect to any
one of the subsystems. As before, we define
\begin{eqnarray}
k_{0}= \textrm{min}
\frac{Tr(Q^{T_{X}}\sigma)}{Tr((I-P)\sigma)}\label{k0}
\end{eqnarray}
where the minimum is taken over all separable states $\sigma$.\vskip0.2cm

\noindent \textbf{Result-3:} Given a state $\rho$ whose partial
transposition is negative, then one witness that can detect $\rho$
is
\begin{eqnarray}
W= (| e_{-}\rangle \langle e_{-}|)^{T_{A}}
\end{eqnarray}
where $| e_{-}\rangle$ is an eigenvector corresponding to a
negative eigenvalue of $\rho^{T_{A}}$.\vskip0.2cm

\noindent \textbf{Result-4:} For a given pair of entangled states
$\rho_{1}$ and $\rho_{2}$ an EW, $W'$ common to both of them
exists if and only if for $0\leq\lambda\leq
1,\lambda\rho_{1}+(1-\lambda)\rho_{2} $ is an entangled state
\cite{12}.

\section{Common Schmidt number Witness}

Consider the space $H^{N} \otimes H^{M}$, with $N<M$. Define $S_{k}$ to be the 
set of states whose Schmidt number is $\leq k$. Thus, $S_{1}$ is the set of 
separable states and the different states share the relation $S_{1} \subset S_{2} \subset S_{3}.... \subset S_{k}..\subset S_{N}$ and  are convex\cite{10,11}.\\
A $k$ Schmidt witness($kSW$),$W^{S}$ is defined as \cite{10} 
\begin{eqnarray}
Tr(W^{S}\sigma)\geq0, ~~\forall \sigma \in S_{k-1}~~~~ \\
Tr(W^{S}\rho)<0 ~~ for~~ at~~ least~~ one~~ \rho \in S_{k}
\end{eqnarray}
A well-known example of a $kSW$ is $I-\frac{m}{k-1}P$ \cite{10} where $m$ and 
$k$ respectively denote the dimension and Schmidt number and $P$ is a 
projector on $\frac{1}{\sqrt{m}}\Sigma_{i=0}^{m-1}\vert ii\rangle $.\\
\textbf{Proposition-I:} Suppose $ \rho_{1} $ and $ \rho_{2} $ are Schmidt number
 $k$ states. If there exists a common $k$SW for $\rho_{1}$ and $\rho_{2}$, then 
the Schmidt number of their convex combination will also be $k$. In other words
the  Schmidt number of $\lambda \rho_{1}+(1-\lambda) \rho_{2}$ is also $k$ ($\lambda\in[0,1]$).\\
\textbf{Proof:} Since $\rho_{1}$,$\rho_{2}$ are in $S_{k}$ and $S_{k}$ is convex , $\lambda \rho_{1}+(1-\lambda) \rho_{2}$ cannot have a Schmidt number  $>k$. \\
Now, let $W^{S}$ be the common $kSW$ for  $\rho_{1}$ and $\rho_{2}$ . As a result
\begin{eqnarray}
Tr(W^{S}(\lambda \rho_{1}+(1-\lambda) \rho_{2}))
 =\lambda Tr(W^{S}\rho_{1}) + (1-\lambda) Tr(W^{S}\rho_{2})
 < 0
\end{eqnarray}
since $Tr(W^{S}\rho_{1})<0$, $ Tr(W^{S}\rho_{2})<0$.
Thus the Schmidt number of $\lambda \rho_{1}+(1-\lambda) \rho_{2}$ is also $k$.\\\\
\textbf{Proposition-II: } Suppose $\delta_{1}$ and $\delta_{2}$ are two states 
with Schmidt number (SN) $k_{1}$ and $k_{2}$ respectively where $k_{1} > k_{2}$. Then a common witness $W_{k}$, if it exists,  will be of class $k$, where 
$k=min(k_{1},k_{2})$.\\
\textbf{Proof:} It follows from the definition of Schmidt number witness that there exists a $k_{1}SW$, $W_{k_{1}}$ for which $Tr(W_{k_{1}}\delta_{1})<0$, but $Tr(W_{k_{1}}\delta_{2})\geq 0$. Therefore a common witness if it exists should be of class $k$ where $k=min(k_{1},k_{2})$.\\\\
\textbf{Example-I: Convex combination of two pure SN 3 states}\\
Consider the states $\vert \Phi_{1}\rangle = a\vert00\rangle + b\vert 11\rangle + \sqrt{1-a^{2}-b^{2}}\vert 22\rangle$ and $\vert \Phi_{2}\rangle= x\vert00\rangle + y\vert 11\rangle + \sqrt{1-x^{2}-y^{2}}\vert 22\rangle$. A $3SW$ of the form $W^{S3}=I-\frac{3}{2}P$ detects both states for many ranges of $a , b , c, x, y, z$ (one such range is $0.25\leq a \leq0.65,0.25\leq b \leq0.65,0.25\leq x \leq0.65,0.25\leq y \leq0.65$). Therefore, for those ranges, $W^{S3}$ is a common 
witness for the states $\vert \Phi_{1}\rangle$ and $\vert \Phi_{2}\rangle$ and thus their convex combination will have SN 3. ($P$ is a projector on $\frac{1}{\sqrt{3}}\Sigma_{i=0}^{2}\vert ii\rangle $)\\\\
\textbf{Example-II: Convex combination of a pure SN 3 state and a pure SN 2 state}\\
Consider now the state $\vert \Phi_{1}\rangle = a\vert00\rangle + b\vert 11\rangle + \sqrt{1-a^{2}-b^{2}}\vert 22\rangle$ and $\vert \chi \rangle = t\vert 00 \rangle + \sqrt{1-t^{2}}\vert 11\rangle$. Here a $2SW$ of the form $W^{S2}=I-3P$ detects both of them whereas the previous $3SW$ fails to qualify as a common witness.\\\\
\textbf{Example-III: Convex combination of a mixed state and a pure SN 2 state} \\
Consider the two-qutrit isotropic state $\Omega= \alpha P+\frac{1-\alpha}{9}I$ 
with $(-\frac{1}{8}\leq \alpha \leq 1)$. The $2SW$, $W^{S2}$ detects it $\forall 1 \geq \alpha > \frac{1}{4}$, which is exactly the range for which the 
isotropic state is entangled. As a result,  the $2SW$ detects $\lambda \Omega + (1-\lambda) \vert\chi\rangle \langle \chi \vert (\lambda \in [0,1])$.

\section{Methods to construct common entanglement witness}\label{common}

\noindent \textbf{Case-I}: Let us consider that the two states
described by the density operators $\rho_{1}$ and $\rho_{2}$ be
negative partial transpose (NPT) states.
Let us further assume that the two sets $S_{1}$ and $S_{2}$
consist of the set of all eigenvectors of $\rho_{1}^{^{T_{A}}}$
and $\rho_{2}^{^{T_{A}}}$ corresponding to their negative
eigenvalues. In set builder notation, $S_{1}$ and $S_{2}$ can be
expressed as $S_{1}=\{|x\rangle:
\rho_{1}^{^{T_{A}}}|x\rangle=\lambda_{-}|x\rangle,~\lambda_{-}~~
\text{is a negative eigenvalue of}~~ \rho_{1}^{^{T_{A}}} \}$ and
$S_{2}=\{|y\rangle:
\rho_{2}^{^{T_{A}}}|y\rangle=\alpha_{-}|y\rangle,~\alpha_{-}
~~\text{is a negative eigenvalue of}~~ \rho_{2}^{^{T_{A}}} \}$.
 Now we propose the following theorem:\vskip0.1cm
\textbf{Theorem 1:} If $S_{1} \cap S_{2}\neq\phi$, then there
exists a common witness detecting not only $\rho_{1}$ and $\rho_{2}$ both
but also all the states lying on the straight line joining $\rho_{1}$ and $\rho_{2}$.\\
\textbf{Proof:} Let $S_{1} \cap S_{2}\neq\phi$ . Then there exists
a non-zero vector $|\eta\rangle \in S_{1} \cap S_{2}$. Let
$W=(|\eta\rangle\langle \eta |)^{^{T_{A}}}$. This gives
\begin{eqnarray}
Tr(W\rho_{1})= Tr((|\eta\rangle\langle \eta
|)^{^{T_{A}}}\rho_{1})
=Tr((|\eta\rangle\langle \eta |)\rho_{1}^{^{T_{A}}})< 0
\end{eqnarray}
With similar justifications,
\begin{eqnarray}
Tr(W\rho_{2})<0
\end{eqnarray}
If now we consider $\rho=\lambda\rho_{1}+(1-\lambda)\rho_{2},
\lambda \in [0,1]$, then $Tr(W\rho)<0$. Hence the theorem.\\
\textbf{Case-II:} Let $\delta_{1}$ and $\delta_{2}$ be two edge
states. We know that a witness operator of the form
$W_{edge}=P+Q^{T_{A}}-\varepsilon I$ can detect an edge state
$\delta$ if $P$ is a projector on ker($\delta$) and $Q$ a
projector on ker($\delta^{T_{A}}$) and
$0<\varepsilon\leq\varepsilon_{0}=inf_{|e,f\rangle}\langle
e,f|P+Q^{T_{A}}|e,f\rangle$ where $|e,f\rangle$ is a product vector \cite{8}.
Thus we propose:\\
\textbf{Theorem 2:} $W_{edge}$ can detect both $\delta_{1}$ and
$\delta_{2}$ if 
$ \texttt{dim} (\texttt{ker}(\delta_{1}) \cap
\texttt{ker}(\delta_{2}))>0 $ or $ \texttt{dim}
(\texttt{ker}(\delta_{1}^{T_{A}}) \cap
\texttt{ker}(\delta_{2}^{T_{A}}))>0 $. \\
\textbf{Proof: } Let $ \texttt{dim} (\texttt{ker}(\delta_{1}) \cap
\texttt{ker}(\delta_{2}))>0 $, i.e., there exists at least one
non-zero eigenvector $|a\rangle \in \texttt{ker}(\delta_{1}) \cap
\texttt{ker}(\delta_{2})$. We assume $P=|a\rangle \langle a|$.\\
Further, let $\texttt{dim} (\texttt{ker}(\delta_{1}^{T_{A}}) \cap
\texttt{ker}(\delta_{2}^{T_{A}}))>0$. We take $|b\rangle \in
\texttt{ker}(\delta_{1}^{T_{A}}) \cap
\texttt{ker}(\delta_{2}^{T_{A}})$. Assume $Q=(|b\rangle\langle
b|)^{T_{A}}$. On taking $W_{edge}=P+Q^{T_{A}}-\varepsilon I$ with
the above mentioned definition of $\varepsilon$ , we obtain
\begin{eqnarray}
Tr(W_{edge}\delta_{1})<0 ~~~ \text{and} ~~~
Tr(W_{edge}\delta_{2})<0
\end{eqnarray}
Consequently, $W_{edge}$ detects $\delta=\lambda
\delta_{1}+(1-\lambda)\delta_{2}$ for $0\leq\lambda\leq1 $ since
\begin{eqnarray}
Tr(W_{edge}\delta)= Tr(W_{edge}(\lambda
\delta_{1}+(1-\lambda)\delta_{2}))<0
\end{eqnarray}
Thus $W_{edge}$ is a common witness for $\delta_{1}$ and
$\delta_{2}$ and detects any convex combination of $\delta_{1}$
and $\delta_{2}$.\\
\textbf{Case-III:} Let $\delta^{1}_{tri}$ and $\delta^{2}_{tri}$
be two tripartite edge states. Using (\ref{multiwitness}) we have
the following theorem:\\
\textbf{Theorem 3:} The witness $W_{tri}$ can detect both the
tripartite edge states $\delta^{1}_{tri}$ and $\delta^{2}_{tri}$
if $ \texttt{dim} (\texttt{ker}(\delta^{1}_{tri}) \cap
\texttt{ker}(\delta^{2}_{tri}))>0 $ or $ \texttt{dim}
(\texttt{ker}((\delta^{1}_{tri})^{T_{X}}) \cap
\texttt{ker}((\delta^{2}_{tri})^{T_{X}}))>0 $. Here $T_{X}$
represents the transposition with respect to any one of the
subsystems. \\
\textbf{Proof: } Proof is similar to \textbf{Theorem 2}.

\section{Examples from Qutrit systems}

Here, we exemplify the methods to find common entanglement witnesses as laid 
down in section \ref{common}
for the different classifications.\\
\textbf{Example 1:} Let us consider the following states in $C^{3}
\otimes C^{3}$:
$\rho_{1}=|\psi_{1}\rangle \langle \psi_{1}|$ and
$\rho_{2}=|\psi_{2}\rangle \langle \psi_{2}|$, 
where $|\psi_{1}\rangle=\frac{1}{\sqrt{2}}(|00\rangle +
|11\rangle)$ and $|\psi_{2}\rangle=\frac{1}{\sqrt{3}}(|00\rangle +
|11\rangle + |22\rangle)$. 
On observation we find an eigenvector $|e_{-}\rangle=|01\rangle -
|10\rangle$ common to $\rho_{1}^{T_{A}}$ and $\rho_{2}^{T_{A}}$
corresponding to their respective  negative eigenvalues.
On defining $U=|e_{-}\rangle \langle e_{-}|$ and $W=U^{T_{A}}$, we
obtain $Tr(W\rho_{1})<0$ and $Tr(W\rho_{2})<0$.
Therefore, $W$ is a common witness to the entanglement in
$\rho_{1}$ and $\rho_{2}$. 
Hence we can conclude that $\rho=\lambda\rho_{1}+(1-\lambda)\rho_{2}$
is entangled for all $\lambda \in [0,1]$ and can be detected by
$W$.\\
\textbf{Example 2:}  The following family of edge
states in
$C^{2} \otimes C^{4}$ was introduced in \cite{13}.\\
\begin{eqnarray}
\tau(b,s)=\frac{1}{2(2+b+b^{-1})}\left(%
\begin{array}{cccccccc}
  b & 0 & 0 & 0 & 0 & -1 & 0 & 0 \\
  0 & 1 & 0 & 0 & 0 & 0 & -1 & 0 \\
  0 & 0 & b^{-1} & 0 & 0 & 0 & 0 & -1 \\
  0 & 0 & 0 & 1 & s & 0 & 0 & 0 \\
  0 & 0 & 0 & s & 1 & 0 & 0 & 0 \\
  -1 & 0 & 0 & 0 & 0 & b^{-1} & 0 & 0 \\
  0 & -1 & 0 & 0 & 0 & 0 & 1 & 0 \\
  0 & 0 & -1 & 0 & 0 & 0 & 0 & b \\
\end{array}%
\right)
\end{eqnarray}
\begin{eqnarray}
(\tau(b,s))^{T_{A}}=\frac{1}{2(2+b+b^{-1})}\left(%
\begin{array}{cccccccc}
  b & 0 & 0 & 0 & 0 & 0 & 0 & s \\
  0 & 1 & 0 & 0 & -1 & 0 & 0 & 0 \\
  0 & 0 & b^{-1} & 0 & 0 & -1 & 0 & 0 \\
  0 & 0 & 0 & 1 & 0 & 0 & -1 & 0 \\
  0 & -1 & 0 & 0 & 1 & 0 & 0 & 0 \\
  0 & 0 & -1 & 0 & 0 & b^{-1} & 0 & 0 \\
  0 & 0 & 0 & -1 & 0 & 0 & 1 & 0 \\
  s & 0 & 0 & 0 & 0 & 0 & 0 & b \\
\end{array}%
\right)
\end{eqnarray}
where $0<b<1$ and $|s|<b$.
We consider $\delta_{1}=\tau(0.4,0)$ and
$\delta_{2}=\tau(0.5,0)$. It is observed that the eigenvector
$|01\rangle + |12\rangle \in
ker(\delta_{1}) \cap ker(\delta_{2})$. 
Further, the eigenvector $|03\rangle + |12\rangle$ and $|01\rangle
+ |10\rangle$ lies in $ker(\delta_{1}^{T_{A}})\cap
ker(\delta_{2}^{^{^{T_{A}}}})$. 
Taking the projectors as defined in Theorem (2) and $\varepsilon$
as in Result-1, we obtain the witness
\begin{eqnarray}
W_{edge}=\left(%
\begin{array}{cccccccc}
  -\varepsilon & 0 & 0 & 0 & 0 & 1 & 0 & 0 \\
  0 & -\varepsilon+2 & 0 & 0 & 0 & 0 & 1 & 0 \\
  0 & 0 & -\varepsilon & 0 & 0 & 0 & 0 & 1 \\
  0 & 0 & 0 & -\varepsilon+1 & 0 & 0 & 0 & 0 \\
  0 & 0 & 0 & 0 & -\varepsilon+1 & 0 & 0 & 0 \\
  1 & 0 & 0 & 0 & 0 & -\varepsilon & 0 & 0 \\
  0 & 1 & 0 & 0 & 0 & 0 & -\varepsilon+2 & 0 \\
  0 & 0 & 1 & 0 & 0 & 0 & 0 & -\varepsilon \\
\end{array}%
\right)
\end{eqnarray}
This gives $Tr(W_{edge}\delta_{1})<0$ and $Tr(W_{edge}\delta_{2})<0$.
Thus, $W_{edge}$ is a common witness and also detects the class of
states 
$\delta=\lambda\delta_{1}+(1-\lambda)\delta_{2},0\leq\lambda\leq1$.\\
\textbf{Example 3:} We consider the following class of tripartite edge states
as proposed in \cite{15}:\\
\begin{eqnarray}
\delta_{tri}(a,b,c)= \frac{1}{n}
\left(%
\begin{array}{cccccccc}
  1 & 0 & 0 & 0 & 0 & 0 & 0 & 1 \\
  0 & a & 0 & 0 & 0 & 0 & 0 & 0 \\
  0 & 0 & b & 0 & 0 & 0 & 0 & 0 \\
  0 & 0 & 0 & c & 0 & 0 & 0 & 0 \\
  0 & 0 & 0 & 0 & \frac{1}{c} & 0 & 0 & 0 \\
  0 & 0 & 0 & 0 & 0 & \frac{1}{b} & 0 & 0 \\
  0 & 0 & 0 & 0 & 0 & 0 & \frac{1}{a} & 0 \\
  1 & 0 & 0 & 0 & 0 & 0 & 0 & 1 \\
\end{array}%
\right)
\end{eqnarray}
where $n=2+a+b+c+1/a+1/b+1/c$ and the basis is taken in the order
$|000\rangle,|001\rangle,|010\rangle,|011\rangle,\\|100\rangle,
|101\rangle,|110\rangle,|111\rangle$. The partial transpose with
respect to system $C$ is given by
\begin{eqnarray}
\delta_{tri}^{T_{C}}(a,b,c)= \frac{1}{n}
\left(%
\begin{array}{cccccccc}
  1 & 0 & 0 & 0 & 0 & 0 & 0 & 0\\
  0 & a & 0 & 0 & 0 & 0 & 1 & 0 \\
  0 & 0 & b & 0 & 0 & 0 & 0 & 0 \\
  0 & 0 & 0 & c & 0 & 0 & 0 & 0 \\
  0 & 0 & 0 & 0 & \frac{1}{c} & 0 & 0 & 0 \\
  0 & 0 & 0 & 0 & 0 & \frac{1}{b} & 0 & 0 \\
  0 & 1 & 0 & 0 & 0 & 0 & \frac{1}{a} & 0 \\
  0 & 0 & 0 & 0 & 0 & 0 & 0 & 1 \\
\end{array}%
\right)
\end{eqnarray}
Next we take the edge states $\delta^{1}_{tri}=\delta_{tri}(1,1,1)$ and
$\delta^{2}_{tri}=\delta_{tri}(1,2,2)$. It is observed that $|111\rangle-|000\rangle \in
\texttt{ker}(\delta^{1}_{tri}) \cap \texttt{ker}(\delta^{2}_{tri})$ and
$|110\rangle-|001\rangle \in \texttt{ker}((\delta^{1}_{tri})^{T_{C}}) \cap
\texttt{ker}((\delta^{2}_{tri})^{T_{C}})$. Now, taking the projectors and $k$ as defined in
Result 2 the witness is
\begin{eqnarray}
W_{tri}=
\left(%
\begin{array}{cccccccc}
  0 & 0 & 0 & 0 & 0 & 0 & 0 & -k-1\\
  0 & 1-k & 0 & 0 & 0 & 0 & 0 & 0 \\
  0 & 0 & -k & 0 & 0 & 0 & 0 & 0 \\
  0 & 0 & 0 & -k & 0 & 0 & 0 & 0 \\
  0 & 0 & 0 & 0 & -k & 0 & 0 & 0 \\
  0 & 0 & 0 & 0 & 0 & -k & 0 & 0 \\
  0 & 0 & 0 & 0 & 0 & 0 & 1-k & 0 \\
  -k-1 & 0 & 0 & 0 & 0 & 0 & 0 & 0 \\
\end{array}%
\right)
\end{eqnarray}
It is found that $W_{tri}$ detects both $\delta^{1}_{tri}$ and $\delta^{2}_{tri}$ , thus
detecting the states $\delta^{12}_{tri}=\lambda\delta^{1}_{tri}+(1-\lambda)\delta^{2}_{tri},\forall \lambda \in [0,1]$.\\

\section{Common decomposable and non-decomposable witness operators }

Central to the idea of the detection of a PPTES  is a
non-decomposable witness which can successfully identify a PPTES
in contrast to a decomposable witness which fails in this purpose.
If we are given two states described by the density operators
$\Delta_{1}$ and $\Delta_{2}$ then we can construct a witness
operator common not only to the states $\Delta_{1}$ and
$\Delta_{2}$ but also to the states lying on the straight line
joining $\Delta_{1}$ and $\Delta_{2}$. Naturally, the next
question as to whether the common witness operator is
decomposable or non-decomposable. The answer 
lies in the nature of the states $\Delta_{1}$ and
$\Delta_{2}$. The decomposable or non-decomposable nature of the
common witness operator depends on the
PPT or NPT nature of the states
$\Delta_{1}$ and $\Delta_{2}$. Let us suppose that $\Delta_{1}$ and 
$\Delta_{2}$ are two
entangled states. Now if we consider the convex combination of
$\Delta_{1}$ and $\Delta_{2}$, i.e.,
$\Delta=\lambda\Delta_{1}+(1-\lambda)\Delta_{2},~~0\leq\lambda\leq1$,
then the common decomposable witness operator and common
non-decomposable witness operator can be seen as:\vskip0.2cm

\noindent \textbf{Common decomposable witness operator: } If both
$\Delta_{1}$ and $\Delta_{2}$ are NPT then a decomposable operator is enough to qualify as
a common witness.\vskip0.2cm

\noindent \textbf{Common non-decomposable witness operator: } If
either $\Delta_{1}$ or $\Delta_{2}$ or both are PPT then the
common witness operator is non-decomposable. \vskip0.2cm

\noindent Note that if $\Delta_{1}$ is PPT and $\Delta_{2}$ is NPT, or
vice-versa, then the state $\Delta$ may be NPT and it may be
detected by a decomposable witness operator, but such a witness
operator will not be common to $\Delta_{1}$ and $\Delta_{2}$,
because either $\Delta_{1}$ or $\Delta_{2}$ is PPT, and a PPT
entangled state cannot be detected by a decomposable witness
operator. Let us understand the above defined common decomposable
and common non-decomposable witness operators by considering the
following two cases: (i) convex combination of a class of PPT
mixed entangled state and a class of NPT pure entangled state and
(ii) convex combination of a class of PPT mixed entangled state
and a class of NPT mixed entangled state.\vskip0.2cm

\noindent \textbf{Case-I: Convex combination of a class of PPT
mixed entangled state and a class of NPT pure entangled
state}\vskip0.1cm

\noindent Let us consider a class of PPT mixed entangled state
\cite{16}
\begin{eqnarray}
\rho_{1}^{e}=\frac{2}{7}\vert\psi^{+}\rangle\langle\psi^{+}\vert+
\frac{\alpha}{7}\varrho_{+}+\frac{5-\alpha}{7}\varrho_{-},~~~3<\alpha\leq4
\label{be}
\end{eqnarray}
where
$\vert\psi^{+}\rangle=\frac{1}{\sqrt{3}}(\vert00\rangle+\vert11\rangle+\vert22\rangle)$,
$\varrho_{+}=\frac{1}{3}(\vert01\rangle\langle01\vert+\vert12\rangle\langle12\vert+
\vert20\rangle\langle20\vert)$ and
$\varrho_{-}=\frac{1}{3}(\vert10\rangle\langle10\vert+\vert21\rangle\langle21\vert+
\vert02\rangle\langle02\vert)$. Further let us consider a pure entangled 
state which is
described by the density operator
\begin{eqnarray}
\rho_{2}^{e}=\beta|00\rangle\langle 00|+\beta\sqrt{1-\beta^{2}}|00\rangle\langle11\vert+
\beta\sqrt{1-\beta^{2}}|11\rangle\langle00\vert+(1-\beta^{2})\vert11\rangle\langle11\vert
\label{sn2}
\end{eqnarray}
The convex combination of the above two states can be described by
the density operator
\begin{eqnarray}
\rho^{e}=\lambda\rho_{1}^{e}+(1-\lambda)\rho_{2}^{e},0\leq\lambda\leq1
\label{nppt}
\end{eqnarray}
Enumerating the eigenvalues of the partial transpose of the state (\ref{nppt}) it is observed that the state has the following characterization:\\
\begin{center}
\begin{tabular}{|c|c|c|c|c|}
\hline Sl. No. & $\lambda$ & $\alpha$ & $\beta$ & Nature of $\rho^{e}$ \\
\hline 1 & $0\leq\lambda<0.75$ & $3<\alpha\leq3.9$ & $0.07 < \beta\leq0.99$ & $(\rho^{e})^{T_{B}}<0$ \\
\hline 2 & $0.75\leq\lambda\leq1$ & $3<\alpha\leq3.9$ & $0\leq\beta\leq0.01$ &  $(\rho^{e})^{T_{B}}\geq0$\\
\hline
\end{tabular}
\end{center}\vskip0.5cm

\noindent Since the state $\rho^{e}$ is free entangled for the
range of three parameters $0\leq\lambda<0.75$, $3<\alpha\leq 3.9$,
$0.07 < \beta\leq0.99$, so a decomposable witness operator is
sufficient to detect it and it is given by
\begin{eqnarray}
W^{d}=(\vert\chi\rangle\langle\chi\vert)^{T_{B}}
\end{eqnarray}
where $\vert\chi\rangle$ is an eigenvector corresponding to a
negative eigenvalue of the state $(\rho^{e})^{T_{B}}$. The witness
operator $W^{d}$ detects $\rho^{e}$ as well as the state
$\rho_{2}^{e}$, but it fails to detect $\rho_{1}^{e}$, as
$\rho_{1}^{e}$ is PPT and $W^{d}$ is decomposable. So, in this case
we are not able to construct a common decomposable witness operator.
However, using Result-1, we can construct a
non-decomposable witness operator in the form
\begin{eqnarray}
W^{nd}=\vert\phi\rangle\langle\phi\vert-\varepsilon I
\end{eqnarray}
where $\vert\phi\rangle\in ker(\rho^{e})$. With this selection, we obtain
\begin{eqnarray}
Tr(W^{nd}\rho^{e})=-\varepsilon<0 ,~~
Tr(W^{nd}\rho_{1}^{e})=-\varepsilon<0 ,~~
Tr(W^{nd}\rho_{2}^{e})=-\varepsilon<0
\end{eqnarray}
The above non-decomposable witness operator $W^{nd}$ not only
detects $\rho_{2}^{e}$ but also detects $\rho_{1}^{e}$, and thus, it
is a common non-decomposable witness operator. Let us now consider the case when
$(\rho^{e})^{T_{B}}\geq 0$ for $0.75\leq\lambda\leq 1$,
$3<\alpha\leq 3.9$, $0\leq\beta\leq 0.01$.  As $\beta\rightarrow0$, the 
state $\rho_{2}^{e}$
approaches the separable projector $\vert11\rangle\langle11\vert$.
Consequently, the convex combination of $\rho_{1}^{e}$ and
$\rho_{2}^{e}$ is PPT. Thus in this scenario, we
can conclude that either all the states lying on the straight line
joining $\rho_{1}^{e}$ and the projector
$\vert11\rangle\langle11\vert$ are separable, or we are incapable
of detecting the most weak bound entangled state.\vskip0.2cm

\noindent \textbf{Case-II: Convex combination of a class of PPT
mixed entangled state and a class of NPT mixed entangled
state}\vskip0.1cm

\noindent Let us consider a class of PPT entangled mixed state and
a class of NPT entangled mixed state which are described by the
density operators
\begin{eqnarray}
\Upsilon_{1}=
\frac{2}{7}\vert\psi^{+}\rangle\langle\psi^{+}\vert+\frac{\alpha}{7}\varrho_{+}+\frac{5-\alpha}{7}\varrho_{-}~~~
 (3 < \alpha\leq 4) \label{be1}
\end{eqnarray}
and
\begin{eqnarray}
\Upsilon_{2}=
\frac{2}{7}\vert\psi^{+}\rangle\langle\psi^{+}\vert+\frac{\gamma}{7}\varrho_{+}+\frac{5-\gamma}{7}\varrho_{-}~~~
  (4 < \gamma\leq 5) \label{npt1}
\end{eqnarray}
respectively.
The convex combination of the states $\Upsilon_{1}$ and
$\Upsilon_{2}$ is given by
\begin{eqnarray}
\Upsilon=\lambda\Upsilon_{1}+(1-\lambda)\Upsilon_{2}~~~ (0\leq\lambda\leq1) \label{nppt1}
\end{eqnarray}
The nature of the resultant state described by the density
operator $\Upsilon$ depends on the values of the mixing parameter
$\lambda$ and the other two parameters $\alpha$ and $\gamma$, as is
given in the table below:
\begin{center}
 \begin{tabular}{|c|c|c|c|c|}
\hline Sl. No. & $\alpha$ & $\gamma$ & $\lambda$ & Nature of $\Upsilon$ \\
\hline 1 & $3<\alpha\leq4$ & $4<\gamma\leq5$ & $0\leq\lambda<\frac{\gamma-4}{\gamma-\alpha}$ & $\Upsilon^{T_{B}}<0$ \\
\hline 2 & $3<\alpha<4$ & $4<\gamma\leq5$ & $\frac{\gamma-4}{\gamma-\alpha}\leq\lambda\leq1$ & $\Upsilon^{T_{B}}\geq0$ \\
\hline 3 & $\alpha=4$ & $4<\gamma\leq5$ & $\lambda=1$ & $\Upsilon^{T_{B}}\geq0$ \\
\hline
\end{tabular}
\end{center}\vskip0.5cm

\noindent The state $\Upsilon$ is NPT for the range of parameters
$3<\alpha\leq4$, $4<\gamma\leq5$,
$0\leq\lambda<\frac{\gamma-4}{\gamma-\alpha}$, and  in this case
the common witness operator is a non-decomposable witness
which detects $\Upsilon$, $\Upsilon_{1}$ and $\Upsilon_{2}$,
whereas a decomposable witness fails to detect all the three
simultaneously. However, in the remaining two cases where the ranges of
three parameters are given by
$3<\alpha<4,4<\gamma\leq5,\frac{\gamma-4}{\gamma-\alpha}\leq\lambda\leq1$
and $\alpha=4,4<\gamma\leq5, \lambda=1$,   we find that the vectors
$|v_{1}\rangle=\vert11\rangle-\vert00\rangle \in ker(\Upsilon_{1})
\cap ker(\Upsilon_{2})$ and
$|v_{2}\rangle=\vert22\rangle-\vert00\rangle \in ker(\Upsilon_{1})
\cap ker(\Upsilon_{2})$. In this scenario, a non-decomposable
witness operator can be constructed, which detects both
$\Upsilon_{1}$, $\Upsilon_{2}$ and hence $\Upsilon$. Such a 
non-decomposable witness operator is of the form
\begin{eqnarray}
\Gamma=P- \varepsilon I
\end{eqnarray}
where $P=|v_{1}\rangle\langle v_{1}|+|v_{2}\rangle\langle v_{2}|$,
and $0<\varepsilon\leq\varepsilon_{0}=inf_{|e,f\rangle}\langle
e,f|P|e,f\rangle$.

\section{Conclusions}

To summarize, in this work we have investigated the conditions for the existence
of common Schmidt number and entanglement witnesses, and proposed methods
for the construction of common witness operators. 
Common entanglement witnesses for pairs of entangled states enable us to detect
a large class of entangled states, {\it viz.},  when a common witness exists 
for two states, it enables us to detect all states lying on the line segment 
joining the two. Certain characteristics of the states help us to construct 
the common witnesses which we have discussed here. We have considered a few
interesting examples of states presented earlier in the literature in the 
context of entanglement witnesses, and these illustrations from qutrit 
systems buttress our claim of suggesting schemes for finding common witnesses. 

Our study shows that the  nature of the common witness is significantly 
dictated by the positivity of the transpose of the two states. Specifically,  
a decomposable witness can never qualify to be a common witness if one the 
states is PPTES. Thus, we demarcate between a common decomposable and 
a nondecomposable  witness. In our analysis of common Schmidt number witnesses 
we find that if the two states are both of SN $k$ and a common SN witness 
exists for them, then the convex combination will be of SN $k$. We conclude by
noting that an interesting question for further study could be to find whether 
the converse of the above statement is true. 

\vskip 0.2in

{\it Acknowledgements:}
We thank  Guruprasad Kar for his useful suggestions. ASM acknowledges support
from the DST Project SR/S2/PU-16/2007.


\begin{thebibliography}{99}
\bibitem{1} C. H. Bennett, G. Brassard, C. Crepeau, R. Jozsa, A. Peres and
W. K. Wootters, Phys. Rev. Lett. \textbf{70}, 1895 (1993).
\bibitem{2} C. H. Bennett and S. J. Wiesner, Phys. Rev. Lett. \textbf{69},
2881 (1992).
\bibitem{3} C. H. Bennett and G. Brassard, Proc. of IEEE Int.
Conf. on Comp., Sys., and Sig. Proc., Bangalore, India, 175
(1984).
\bibitem{4} A. Barenco, D. Deutsch, A. Ekert and R. Jozsa, Phys. Rev.
Lett. \textbf{74}, 4083 (1995).
\bibitem{5} A. Peres, Phys. Rev. Lett. \textbf{77}, 1413 (1996).
\bibitem{6} M. Horodecki, P. Horodecki and R. Horodecki, Phys. Lett. A
\textbf{223}, 1 (1996).
\bibitem{7} B. Terhal, Linear Algebr. Appl. {\bf 323}, 61 (2001).
\bibitem{8} M. Lewenstein, B. Krauss, J. I. Cirac and P. Horodecki, Phys. Rev. A \textbf{62}, 052310
(2000).
\bibitem{9}A. C. Doherty , P. A. Parrilo and F. M. Spedalieri, Phys. Rev. A \textbf{69},
022308 (2004).
\bibitem{10} A. Sanpera, D. Bru$\beta$ and M. Lewenstein , Phys. Rev . A \textbf{63}, 050301(R) (2001).
\bibitem{11} B. M. Terhal and P. Horodecki, Phys. Rev . A \textbf{61}, 
040301(R) (2000).
\bibitem{12} Y. C. Wu and G. C. Guo, Phys. Rev. A \textbf{75}, 052333 (2007).
\bibitem{13} R. Augusiak, J. Grabowski, M . Kus and
M. Lewenstein, Optics Communications \textbf{283}, 805 (2010).
\bibitem{14} N. Ganguly and S. Adhikari, Phys. Rev. A \textbf{80},
032331 (2009).
\bibitem{15} A. Acin, D. Bru$\beta$, M. Lewenstein and
A. Sanpera, Phys. Rev. Lett \textbf{87}, 040401 (2001).
\bibitem{16} P. Horodecki, M. Horodecki, R. Horodecki, Phys. Rev. Lett. \textbf{82}, 1056 (1999).
\end{thebibliography}
\end{document}